\documentstyle[aps,prb,multicol,epsf]{revtex}
\begin{document}
\draft
\title{The phase diagram of high-$T_c$'s: Influence of anisotropy and disorder}
\author{E. A. Jagla and C. A. Balseiro}
\address{Comisi\'{o}n Nacional de Energ\'{\i}a At\'{o}mica,\\
Centro At\'{o}mico Bariloche and Instituto Balseiro\\
8400 San Carlos de Bariloche, Argentina}
\maketitle

\begin{abstract}
We propose a phase diagram for the vortex structure of high 
temperature superconductors which incorporates the effects of anisotropy and
disorder. It is based on numerical simulations using the three-dimensional 
Josephson junction array model. We support the results with  an 
estimation of the internal energy and
configurational entropy of the system. Our results
give a unified picture of the behavior of the vortex lattice, covering from the very 
anysotropic Bi$_2$Sr$_2$CaCu$_2$O$_8$ to the less anisotropic 
YBa$_2$Cu$_3$O$_7$, 
and from the first order 
melting ocurring in clean samples to the continuous transitions observed in samples with defects.
\end{abstract}
                   
                   
\begin{multicols}{2}

\section{Introduction}

The phase diagram of high temperature superconductors in the mixed state has
provided an astonishingly broad field to workers in -among other fields-
many body problems, polymer physics, low dimensional systems, critical
phenomena, and statistical physics in general. The main reason of this
situation is the great number of parameters that define the behavior of the
vortex structure. On the other hand, the same abundance of parameters
defining the system turns difficult to find a unified description of all
features observed in experiments. Some of the main parameters that define
the behavior of the vortex structure are the external magnetic field $H$,
temperature $T$, anisotropy $\eta $, and the disorder (which produces a
non-homogeneous pinning potential for the vortices) that at this moment we
loosely characterize by a parameter $D$. There are convincing explanations
of the main characteristics of different sectors of this multi-dimensional
phase diagram, such as the first order melting of the vortex-lattice in
clean samples, the continuous melting of a glassy phase in disordered
samples, or the existence of two different superconducting transitions
(perpendicular and parallel to $H$) in some cases (for a review see
Ref. 1). However a
unified, consistent with experiments description of the problem, even in a
qualitative level, is still lacking.

In this paper we propose a qualitative $H$-$T$-$D$-$\eta $ phase diagram of
high-$T_c^{\prime }$s, that reproduces most of the available experimental
results. Our approach is twofold: we use numerical simulation on the
three-dimensional Josephson junction array model to study the behavior of the
system as a function of $D$ and $\eta $, and show that the dependence on $H$
can be deduced from a rescaling of $D$ and $\eta $. The obtained phase
diagram is then rederived using a phenomenological
estimation of the free energy $F$ of the system for different values of $D$
and $\eta $. This estimation relies on the existence of two characteristic
lengths $\xi _c$ and $\xi _{ab}$ parallel and perpendicular to the applied
field $H$ which are supposed to govern the behavior of the system. The
minimizing of $F$ with respect to $\xi _c$ and $\xi _{ab}$ allows one to obtain
the $\xi _c(T)$ and $\xi _{ab}(T)$ functions, which in turn are used to
detect the superconducting transitions.

The reminder of the paper is organized as follows. In the next section we
present the results of the numerical simulations, and discuss the $D$-$\eta $
phase diagram emerging from them. In Sec. III this phase diagram is
qualitatively re-obtained using a proposal for the free energy of the
system. In Sec. IV we indicate that a change in the external magnetic
field can be interpreted as a movements in the $D$-$\eta $ plane, and so the 
$H$-$T$ phase diagram for samples with different $D$ and $\eta $ can be
obtained from the results of the previous sections. We also compare our
results with those found in experimental studies. Finally in Sec. V, we
summarize and conclude.

\section{Numerical simulations}

\subsection{The model}

Our numerical results are based on simulations performed on the three
dimensional (3D) Josephson junction array (JJA) model on a stacked
triangular network. Each junction is modelled by an ideal Josephson junction with
critical current $I_c$ shunted by a normal resistance $R$ and its attached
Johnson noise generator, which accounts for the effects of temperature. An
external {\it c} directed magnetic field is included. The variables that
characterize the model are the phase of a superconducting order parameter
defined on the nodes of the lattice. Vortices form in the system as
singularities in the distribution of these phases. The details of the model
have been discussed elsewhere.\cite{mingo2,jb1} The 3D JJA model has been
previously used to show the first order melting of the vortex lattice in
clean systems.\cite{huse,mingo1} Both thermodynamical and transport
signatures of this first order transition were obtained, in close relation
to experimental results.\cite{hugo} In addition, using the same model we
have shown that disorder can destroy the first order transition.\cite{jb4}

We study here the model further by systematically exploring the case of
anisotropic and disordered samples. We introduce anisotropy by reducing the
critical current of the {\it c} axis directed junctions by a factor $\eta ^2$%
, and at the same time increasing the {\it c} axis normal resistance by the
same factor. Disorder is introduced by randomly varying the critical current
of the junctions through the lattice. As vortices gain energy when close to
a low critical current region, the effect of randomizing the critical
currents is to provide a nearly random pinning potential for vortices. We
characterize the disorder by a parameter $D$ which is defined as $D\equiv
\left( I_c^{\max }-I_c^{\min }\right) /\left( I_c^{\max }+I_c^{\min }\right) 
$, where $I_c^{\max }$ and $I_c^{\min }$ are the maximum and minimum value
of the critical current of the junctions through the sample. The probability
distribution between $I_c^{\max }$ and $I_c^{\min }$ is taken flat.

We carried out simulations for $H=1/6$ flux quanta per plaquette. This is
the value used in Refs. 4 and 5. It produces a ground state
(for a clean sample) which is commensurate with the subjacent triangular
lattice, so no frustration effects are expected. Although the value $1/6$ is
rather large and effects of the substrate may be observable, we expect the
physics of the problem to be qualitatively well described. In particular, we
assume that for a clean sample the first order melting observed in
simulations is in fact the counterpart of the experimental observations.\cite{hugo,hernan,zeldov,lopez1,lopez2} It would be
interesting to perform simulations at lower (commensurate) fields, such as $%
1/14$ or $1/36$. However, the sample size needed to minimize size effects
make the computing time be exceedingly large.

\subsection{Results}

All simulations presented here were performed for $H=1/6$, with boundary
conditions as in Ref. 7, and for a sample of $%
L_{ab}\times L_{ab}\times L_c=18\times 18\times 18$ junctions. We
characterize the superconducting transitions by measuring the resistivity of
the sample when a small current (typically 1/100 of the mean critical
current in the corresponding direction) is applied along the {\it ab} or 
{\it c} direction. We will observe two well different behaviors of the
resistivities as a function of temperature: in some cases resistivities have
a jump from zero to a finite value at a given temperature. This jump in the
resistivity corresponds (see the discussions in Sec. IV) to a first order
phase transition of the vortex lattice. In some other cases we will obtain
that resistivities as a function of temperature smoothly depart from zero.
We will refer to this behavior as a {\em continuous transition}. We do not
claim at this point about whether these continuous transitions are or are
not real phase transitions, they can be crossovers as well. In Sec. V we
present a discussion on this point.

\begin{figure}
\epsfxsize=8.7truecm
\vbox{\epsffile{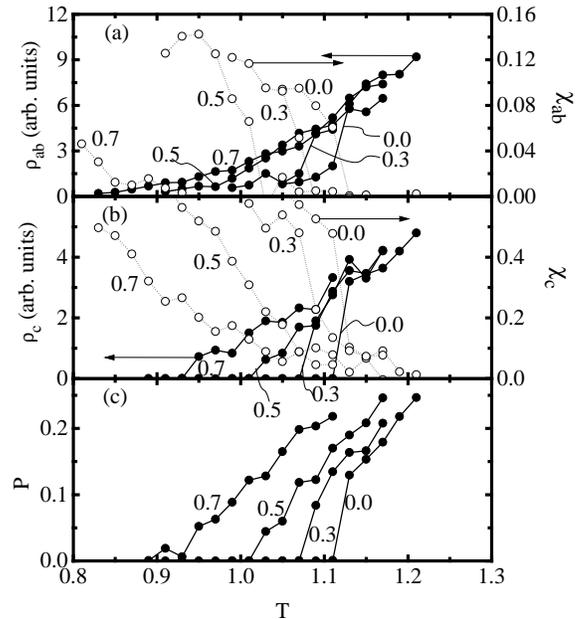}}
\narrowtext
\caption{Resistivities $\rho _{ab},$ $\rho _c,$ helicity modulus $\chi
_{ab}$ and $\chi _{c},$ and percolation probability $P$ as a
function of temperature for $\eta =1$ and different values of the disorder $%
D $ as indicated. Temperature is
measured in units of the mean Josephson energy of the in-plane junctions.
Magnetic field is $H=1/6$ quantum fluxes per plaquette.}
\label{uno}
\end{figure}

We start by showing in Fig. \ref{uno}(a) and (b) (full symbols) the
results for the {\it ab} plane and {\it c} axis resistivity of an isotropic
sample as a function of temperature for different values of the disorder
$D$. For $D=0$ we obtain a jump in the
resistivities at a well defined temperature (the low temperature tails in $%
\rho _{ab}$ are due to surface effects). This temperature is the (first
order) melting temperature $T_m$ of the system. At $T_m$ the superconducting
coherence is lost discontinuously in all directions. In other words,
vortices passes from a solid phase to an entangled liquid phase. This
situation persists for low values of disorder. When disorder increases
further ($D>0.3$ in our simulations), the first order transition is lost,
and two continuous transitions at different temperatures $T_i$ -for $\rho
_{ab}$- and $T_p$ -for $\rho _c$- are obtained. Superconducting coherence
along the {\it ab} plane is lost at $T_i$ when increasing the temperature,
but the system is still superconducting along the {\it c} direction. At the
higher temperature $T_p$ the parallel to field superconducting coherence is
lost. For $T_i<T<T_p$ the vortex structure corresponds to that of a
disentangled vortex liquid.

Thermodynamical measurements agree with this picture. We calculated the
helicity modulus parallel ($\chi _{c}$) and perpendicular ($\chi
_{ab}$) to the field. Helicity modulus measures the influence on the
energy of the system of a twist in the boundary conditions, and has to be
different from zero to indicate superconducting coherence.\cite{helmod} Open
symbols in Fig. \ref{uno}(a) and (b) show the values $\chi $ that correspond
to the resistivity curves. In the case of the first order transition both $%
\chi _{c}$ and $\chi _{ab}$ have an abrupt drop at the melting
temperature. For the highly disordered case $\chi _{ab}$ has a (smoother)
decrease at $T_i$, whereas $\chi _{c}$ becomes nearly zero at $T_p$.
The transition at $T_p$ is a percolation phase transition of the vortex
structure as discussed in Refs. 14 and 12, that can be characterized by a
percolation probability $P.$ The value of $P$ is the probability that a
vortex path traversing the sample perpendicularly to the applied field
exists.
This value is zero below the transition
and one above, with a transition zone that becomes narrower when $L_{ab}$
increases. The plot of $P$ vs temperature is shown in Fig. \ref{uno}(c). We see that $\rho _c$ is different from (equal to) zero if $P$ is
different from (equal to) zero.

\begin{figure}
\epsfxsize=8.7truecm
\vbox{\epsffile{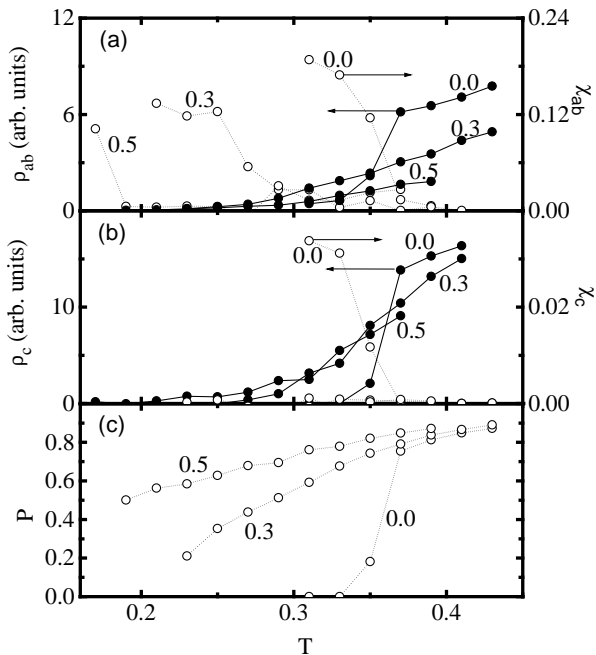}}
\narrowtext
\caption{Same as Fig. 1 but for a sample with anisotropy $\eta ^2=20$.
The values of $\chi_{c}$ for $D=0.3$ and $D=0.5$ are nearly zero within
the considered temperature range.}
\label{dos}
\end{figure}

In Fig. \ref{dos} we give the corresponding results for anisotropy $\eta
^2=20$. The first order melting transition occurring for low disorder is
similar to that observed for $\eta =1$, i.e., the inter-plane coherence is
lost at the same temperature than the in-plane coherence. However, for
higher values of disorder some differences occur: the inter-plane resistive
transition occurs (to our numerical precision) at the same temperature as
the in-plane transition $T_i$, but the percolation transition temperature $%
T_p$ (which for low anisotropy coincides with the resistive inter-plane
transition) occurs for lower temperatures, i.e., $T_p<T_i$. This indicates
that for high anisotropy, there is a temperature range $T_p<T<T_i$ for which
percolation paths across the samples exist, however these path are not
mobile (presumably because they are pinned to the planes, as $T<T_i$) and
dissipation is not observed.\cite{nota1}

From the above discussion on Figs. \ref{uno} and \ref{dos}, the following
picture emerges: at low disorder the vortex lattice melts through a first
order phase transition. When disorder increases this transition breaks into
two continuous ones, where in-plane and inter-plane coherence are lost at
different temperatures $T_i$ and $T_p$. Whether $T_i$ is larger or lower
than $T_p$ depends on the anisotropy of the system. For nearly isotropic
samples $T_i$ is lower than $T_p$, and a disentangled vortex liquid phase
exists for $T_i<T<T_p$. For highly anisotropic samples $T_p$ is lower than $%
T_i$ although dissipation along the {\it c} axis is observed only for $T>T_i$%
.

\begin{figure}
\epsfxsize=8.7truecm
\vbox{\epsffile{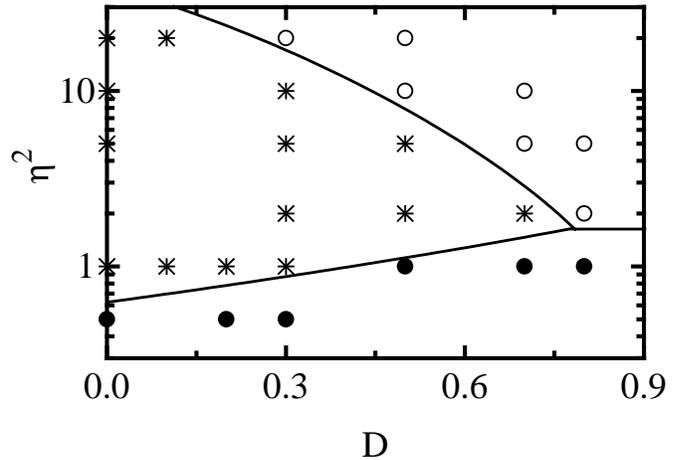}}
\narrowtext
\caption{Disorder-anisotropy phase diagram as obtained from the simulations.
Stars indicate points where the melting transition is first order. Full
circles are points where $T_i<T_p$, and hollow circles are points where $%
T_p<T_i$. The continuous line is a sketch of the frontiers between the
different zones.}
\label{tres}
\end{figure}

In Fig. \ref{tres} we show the numerical $D$-$\eta $ phase diagram as
obtained from simulations. Stars indicate points where the melting
transition is first order. Full circles are points where $T_i<T_p$, and
hollow circles are points where $T_p<T_i$. This diagram shows the three
different zones discussed above. The continuous line is a sketch of the
frontiers between the different zones. In the next section we show that the
main characteristics of this phase diagram can be derived from a simplified
description of the vortex structure.

\section{A simplified model}

We saw in the previous section that a single first order transition is
observed at low disorder, whereas two continuous ones are obtained in
disordered samples. These features suggest that the system can be described
as having in general two different transitions, but that in certain cases
they can merge onto a single first order transition due to some kind of
`interaction'. Here we show that this idea can be formulated more precisely
by estimating the free energy of the system.

\subsection{Free energy functional}

To make this estimation we will consider first a single clean plane. We
suppose that the thermodynamics of that plane can be phenomenologically
described by a quantity $\xi _{ab}$, which is a correlation length: for
distances shorter than $\xi _{ab}$ the system has superconducting coherence,
whereas this coherence is lost for distances larger than $\xi _{ab}$. The
free energy functional ${\cal F}^{2D}$ of the plane has the form 
\begin{equation}
{\cal F}^{2D}=E^{2D}(\xi _{ab})-TS^{2D}(\xi _{ab})  \label{f1}
\end{equation}
(the thermodynamical free energy $F$ is obtained by minimizing with respect
to $\xi _{ab}$). For a system of $L_c$ completely decoupled planes, we would
have 
\begin{equation}
{\cal F}_{\eta \rightarrow \infty }^{3D}=L_cE^{2D}(\xi _{ab})-TL_cS^{2D}(\xi
_{ab})  \label{f2}
\end{equation}
On the other hand, if the coupling between planes is infinite the vortices
are rigid lines and we get 
\begin{equation}
{\cal F}_{\eta \rightarrow 0}^{3D}=-\left( \alpha /\eta ^2\right)
L_c+L_cE^{2D}(\xi _{ab})-TS^{2D}(\xi _{ab}).  \label{f3}
\end{equation}
Note that in this case the entropy term does not have the factor $L_c$
because giving the position of the vortices on one plane automatically
determines the position of vortices in all other planes. The term $-\left(
\alpha /\eta ^2\right) L_c$ accounts for the energy gain due to the coupling
of the planes, $\alpha $ being a numerical constant and $\eta $ the
anisotropy parameter defined before. In an intermediate situation ($0<\eta
<\infty $) the system can be thought as formed by $L_c/\xi _c$ layers
($\xi _c$ is a number that satisfies $1<\xi_c<L_c $). Within each layer the vortices are almost straight lines,
whereas correlation is small between different layers. Within this picture
the free energy of the system is

\end{multicols}
\widetext

\begin{equation}
{\cal F}^{3D}=-\left( \alpha /\eta ^2\right) \left( L_c-L_c/\xi _c\right)
+L_cE^{2D}(\xi _{ab})-T\left( L_c/\xi _c\right) S^{2D}(\xi _{ab}).
\label{f4}
\end{equation} 


\noindent In the first term, $L_c-L_c/\xi _c$ is the number of sites along
the {\it c}-direction at which the system gains an energy $\alpha /\eta ^2$.
The length $\xi _c$ is a correlation length along the {\it c-}direction. For
distances shorter than $\xi _c$ the system possesses superconducting
coherence, whereas this coherence is lost at distances greater than $\xi _c$.

The previous estimation of the free energy of the system is too crude. In
particular, in the form given by Eq. (\ref{f4}), it leads to some
unphysical results. We must modify Eq. (\ref{f4}) slightly in order to
obtain the correct behavior in some limiting cases. However, we will keep a
fundamental property of Eq. (\ref{f4}) and make the guess that the entropy
can be written for the real system as 
\begin{equation}
S=f(\xi _c)S^{2D}(\xi _{ab}),  \label{entropia}
\end{equation}
i.e., as a product of independent functions of $\xi _c$ and $\xi _{ab}$,
with $f(\xi _c)$ and $S^{2D}(\xi _{ab})$ two yet unknown functions. The
basic assumption contained in (\ref{entropia}) is the following: if the
value of $\xi _c$ is kept fixed, then the system behaves as a two
dimensional system with a renormalized temperature. Although this assumption
cannot be fully justified a priori, it is a natural starting point, and
gives sensible results as we will show soon.

We still have to add a term to the free energy which accounts for the effect
of impurities. Impurities decrease the energy of the system when vortices pin
to them. If pinning is uncorrelated the energy gain due to pinning becomes
lower when the vortex positions are more correlated. So we add a term to 
the free energy that increases the energy of the
system when $\xi _{ab}$ and $\xi _c$ increase. The most simple term of this
type is of the form $D\xi _{ab}\xi _c$. Finally, the free energy functional
of the system ${\cal F}\left( \xi _{ab},\xi _c\right) $ (dropping an
irrelevant constant term) is


\begin{equation}
{\cal F}\left( \xi _{ab},\xi _c\right) =\left( \alpha /\eta ^2\right)
L_c/\xi _c+L_cE^{2D}(\xi _{ab})-Tf(\xi _c)S^{2D}(\xi _{ab})+D\xi _{ab}\xi _c.
\label{f5}
\end{equation}


The true expressions for the functions $f$, $E^{2D}$ and $S^{2D}$ are
difficult to establish. However, it is not our aim to give a complete
quantitative description of the free energy of the system but only a
qualitative description of the phase diagram. We will only ask the functions 
$f$, $E^{2D}$ and $S^{2D}$ to reproduce some known limiting cases: if $\xi
_c $ ($\xi _{ab}$) is kept fix we expect the value of $\xi _{ab}(T)$ ($\xi
_c(T) $) obtained by minimizing ${\cal F}$, to be smoothly dependent on
temperature. This corresponds to the absence of first order transitions in
the dynamics of a single plane (a single vortex line).\cite{stroud} However,
when ${\cal F}$ is minimized with respect to {\em both} $\xi _c$ and $\xi
_{ab}$ a {\em \ discontinuity }in $\xi _c(T)$ and $\xi _{ab}(T)$ can appear,
as we show below.

Just to give an example we use for the function $f$ the form $f=\gamma \ln
\left( L_c/\xi _c\right) +1$. The term $\gamma \ln \left( L_c/\xi _c\right) $
($\gamma $ is a numerical constant) is proportional to the entropy of a
single (isolated) vortex line. The constant added assures that the
two-dimensional limiting case is reobtained when $L_c=\xi _c$. In addition,
we will take for the functions $E^{2D}$ and $S^{2D}$ the form $%
E^{2D}=L_{ab}/\xi _{ab}$ and $S^{2D}=\gamma \ln \left( L_{ab}/\xi
_{ab}\right) +1$ , in such a way that we obtain a form for the free energy
functional that is (unphysically!) symmetric (for $\alpha /\eta ^2=1$)
between {\it ab}- and {\it c}-directions. We have tested other forms of the
functions $f$, $E^{2D}$ and $S^{2D}$ (giving the same limiting behavior
discussed above) and found results qualitatively similar to those shown here.

We arrive to our final working expression of the free energy functional (we
will measure $\xi _{ab}$ and $\xi _c$ in units of $L_{ab}$ and $L_c$,
respectively, take $\alpha =1$ by rescaling $\eta ,$ and also rescale $D$)


\begin{equation}
{\cal F}\left( \xi _{ab},\xi _c\right) =1/\eta ^2\xi _c+1/\xi _{ab}-T\left[
\gamma \ln \left( 1/\xi _c\right) +1\right] \left[ \gamma \ln \left( 1/\xi
_{ab}\right) +1\right] +D\xi _{ab}\xi _c  \label{f6}
\end{equation}

\begin{multicols}{2}

The thermodynamical free energy $F$ is obtained by minimizing with respect
to $\xi _{ab}$ and $\xi _c$: 
\begin{equation}
F(T)=\min_{0\leq \xi _{ab}\leq 1}\min_{0\leq \xi _c\leq 1}{\cal F}\left( \xi
_{ab},\xi _c\right)  \label{f7}
\end{equation}

We stress again we do not claim expression (\ref{f6}) is a good detailed
description of the free energy of the system, but only a expression (which
main characteristic is given by Eq. (\ref{entropia})), that will help us to
understand different sectors of the $D$-$\eta $ phase diagram of high-$T_c$%
's.

\subsection{Results}

We present now the results obtained by minimizing the free energy functional
given by expression (\ref{f6}). When doing this minimization we obtain the
free energy $F$ and the lengths $\xi _{ab}$ and $\xi _c$ as a function of
temperature. A first order transition is identified as a discontinuous
derivative of $F$, or equivalently a jump in the values of $\xi _{ab}$ and $%
\xi _c$. When no jump in $\xi _{ab}$ and $\xi _c$ is obtained, the
dependence of $\xi _{ab}$ and $\xi _c$ with temperature gives a clue on how
the superconducting coherence is lost when raising temperature. However, in
this case the identification of a phase transition is not simple, and we can
only identify temperature ranges where coherence along {\it c-} or {\it ab-}%
directions is high or low.

\begin{figure}
\epsfxsize=8.7truecm
\vbox{\epsffile{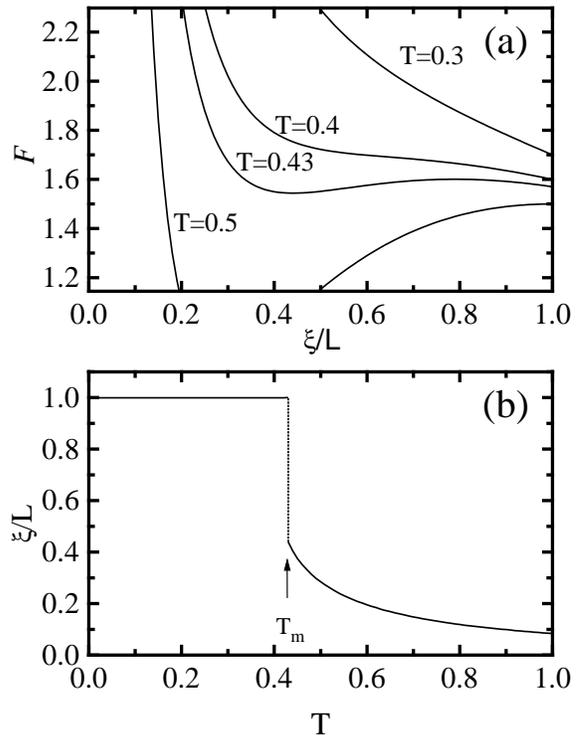}}
\narrowtext
\caption{(a) Free energy functional as a function of correlation length $\xi 
$ for different temperatures with $D=0,$ and $\gamma =2$.(b) Correlation
length as a function of temperature obtained as the minima of the curves in
(a).}
\label{cuatro}
\end{figure}

Let us start with the case $D=0$ and $\eta =1$ (note that we are using a
renormalized anisotropy parameter, so $\eta =1$ does not imply necessarily
an isotropic system). In this case Eq. (\ref{f6}) is symmetric in $\xi
_{ab}$ and $\xi _c$, and in fact the minima of ${\cal F}$ are on the line $%
\xi _{ab}=\xi _c\equiv \xi $, so in Fig. \ref{cuatro} we plot the function 
${\cal F}(\xi ,\xi )$ for different values of the temperature. We see that
when $T \rightarrow 0$ the minimum free energy state corresponds to the
maximum possible
value of $\xi $, i.e., the system is in the ordered state. When temperature
increases a {\em first order} transition to a disordered state occurs. This
is also seen from the behavior of $\xi $ as a function of temperature, as
depicted in Fig. \ref{cuatro}(b). Thus we see that the coupling of two
continuous transition (through the entropy term in (\ref{f6})) can merge them
onto a single one, which is first order. In fact, such sort of mechanism has
been previously proposed to occur in other cases, such as two dimensional
melting, in which the continuous dislocation-unbinding and
disclination-unbinding transitions of the
Kosterlitsz-Thouless-Halperin-Nelson-Young melting theory\cite
{2dmelting1,2dmelting2} can collapse onto a first order melting transition.%
\cite{2dmelting2,2dmelting3}

\begin{figure}
\epsfxsize=8.7truecm
\vbox{\epsffile{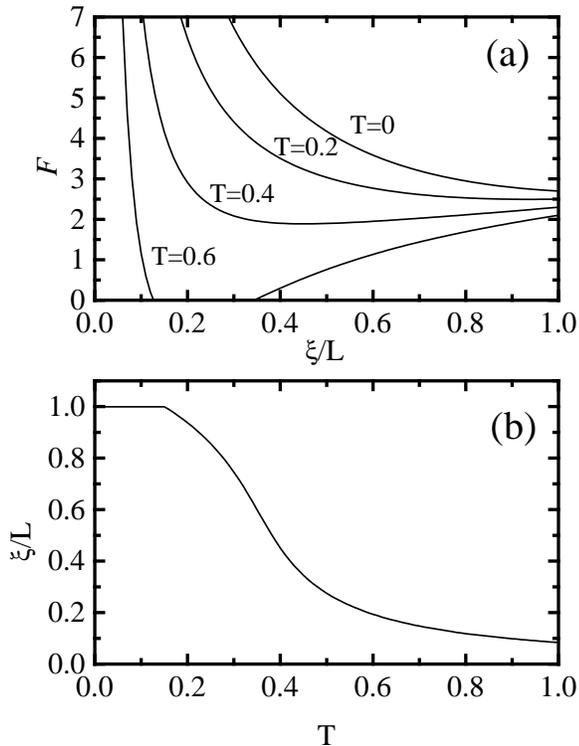}}
\narrowtext
\caption{Same as Fig. 4 but with $D=0.7.$}
\label{cinco}
\end{figure}

\begin{figure}
\epsfxsize=8.7truecm
\vbox{\epsffile{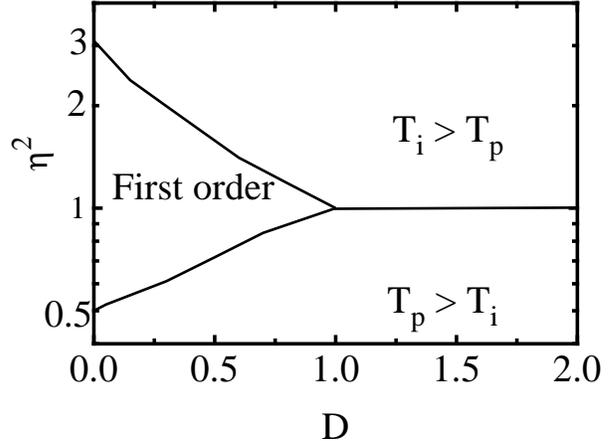}}
\narrowtext
\caption{Disorder-anisotropy phase diagram as obtained by minimizing the
free energy functional (see text for explanation).}
\label{seis}
\end{figure}

If disorder is present in the system it will tend to destroy the melting
transition. In Fig. \ref{cinco} we show results as those of Fig. \ref{cuatro}
but for a value of $D=0.7$. As we see the jumps in $\xi _{ab}$ and $\xi _c$
have disappeared, indicating that the transition is not first order. If the
anisotropy had been chosen different from one, then the temperature at which 
$\xi _{ab}$ and $\xi _c$ take a given value would have been different.
Although we cannot characterize from our simplified model a phase transition
when $\xi _{ab}$ and $\xi _c$ are continuous functions of temperature, it is
tempting to say that if $\xi _{ab}$ ($\xi _c$) drops to zero at lower
temperature than $\xi _c$ ($\xi _{ab}$), then we are in a sector of the
phase diagram where $T_i<T_p$ ($T_p<T_i$). In this way we generate the phase
diagram depicted in Fig. \ref{seis}. As indicated, it is qualitatively
similar to the one obtained from the numerical simulation, and it gives
support{\it \ a posteriori }to the proposal of an entropy of the system of
the form given by Eq. (\ref{entropia}).

Some characteristics of this phase diagram can be analyzed in simple terms.
For example, when $\eta \rightarrow \infty ,$ the system is a set of
decoupled planes, and no first order transition is obtained for any value of 
$D.$ When $\eta \rightarrow 0$ vortices are rigid lines and in fact
effectively two-dimensional, so a first order transition is not obtained in
this case either. This limiting cases give some insight on the form of the
border between first-order and continuous transitions in Figs. \ref{tres}
and \ref{seis}. There is an optimum value of the anisotropy, at which the
first order transition persists up to a highest value of disorder. This
optimum value depends on the thickness of the sample. In fact, as we
discussed in a previous work\cite{jb3} the temperature $T_p$ logaritmically 
{\em decreases }when the thickness of the sample increases. This means that
in Fig. \ref{tres} the border between the zones with $T_i>T_p$ and $T_p>T_i$
moves to lower values of $\eta $. It is thus likely that the optimum value
of $\eta $ for the occurrence of the first order transition also decreases
with sample thickness .

\section{Magnetic field dependence, and comparison with experiments}

Having discussed the $D$-$\eta $ phase diagram for a fixed magnetic field $H$%
, we turn now to the discussion of the dependence on $H$. From the numerical
point of view the direct approach would be to do simulations at different
fields. However, as we discussed above, to reduce the magnetic field to
other commensurate values would require exceedingly large computing time.
Fortunately, there are arguments that suggest that a change of $H$ can be
mapped onto a change of $D$ and $\eta $.

The scaling combination between magnetic field and anisotropy has been given
by Chen and Teitel.\cite{chentei} We generalize here the argument to include
the disorder parameter $D$. Our dimensionless temperature (measured in units
of the mean Josephson energy of the in-plane junctions) can be only a
function of the other dimensionless parameters of the system. These are $D$, 
$H$, and $\eta $. These parameters have different dependences on the
coherence length $\xi _0$. If we identify the discretization parameter in
the {\it ab} plane with a distance of the order of the coherence length $\xi
_0,$ then the critical current of the Josephson junctions along the 
{\it c} direction is proportional
to $\xi _0^2,$ so the anisotropy parameter $\eta $ behaves as $\xi _0^{-1}.$
On the other hand, our dimensionless magnetic field $H$ is given in terms of
the real external magnetic field $H_0$ by the expression $H=H_0\xi _0^2/\phi
_0$ ($\phi _0$ is the flux quantum). For the parameter $D$, we note that $D$
is proportional to the amplitude of the pinning potential. A vortex averages
this random function on an area $\sim \xi _0^2.$ Considering the case of
random (uncorrelated) pinning we find that $D$ depends on $\xi _0$ as $D\sim
\xi _0^{-1}$. Since we are ignoring details of the vortex cores, we expect
the $\xi _0$ dependence to cancel out, and our temperature transition to be
only a function of the $\xi _0$-independent quantities $\eta ^2H$ and $D^2H$.

\begin{figure}
\epsfxsize=8.7truecm
\vbox{\epsffile{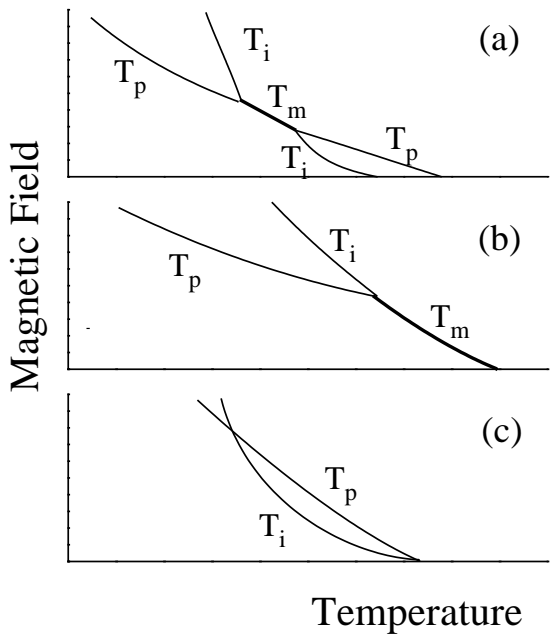}}
\narrowtext
\caption{Qualitative sketch of the $H$-$T$ phase diagram for diferent calues
of $D/\eta .$ (a) General form. (b) $D/\eta $ small. (c) $D/\eta $ high.}
\label{siete}
\end{figure}

We conclude that we can obtain the behavior of the system as a function of
magnetic field from the results of Fig. \ref{tres} on lines with constant $%
D/\eta $. A sketch of the different possibilities is shown in Fig. \ref
{siete}. The general prediction from Fig. \ref{tres} is depicted in Fig. 
\ref{siete}(a). The scales on the axis as well as the extent of the first
order transition depend on the particular value of $D/\eta$. This general
picture has to be modified at very low fields. In fact, a minimum crossover
field\cite{jb2} (given essentially as the field at which the vortex lattice
parameter matches the thickness of the sample) exists, below which $T_i$ and 
$T_p$ are essentially the same (this is because in this case there are so
few vortices in the sample that the transition is enterally due to thermally
generated vortex loops). Different possibilities are depicted in Fig. \ref
{siete} (b), and (c). They correspond to different ranges of values of $%
D/\eta $. Fig. \ref{siete}(b) correspond to $D/\eta $ small, so Fig. \ref
{tres} predicts $T_p<T_i$ at high fields, and a first order zone at
intermediate and low fields. This phase diagram corresponds to that
experimentally obtained for Bi$_2$Sr$_2$CaCu$_2$O,\cite{bismuto} which 
in fact has
the largest value of $\eta ,$ and even to the case of clean
YBa$_2$Cu$_3$O$_7$,%
\cite{ybacuoclean} which has a very low $D$. Fig. \ref{siete}(c) shows the
expected phase diagram for samples with a high value of $D/\eta $. At low
fields $T_i$ is lower than $T_p,$ whereas at high fields a cross to a zone
with $T_i>T_p$ is possible. No first order zone is shown because a curve
defined by a high value of $D/\eta $ in Fig. \ref{tres} do not pass
through the first order zone. The low field part of this phase diagram
corresponds to the one obtained for YBa$_2$Cu$_3$O$_7$ samples with
defects.\cite
{ybacuo} The crossover to a case with $T_i<T_p$ has not been observed,
presumably because of the high fields needed.   

As we indicated above, when $T_i<T_p$ superconducting coherence is lost
along the {\it ab-}plane at lower temperatures than along the {\it c}-axis.
For $T_i<T<T_p$ finite resistance within the planes and zero resistance in
the {\it c}-direction is observed. When $T_i>T_p$ and in the case that $%
T_i>T>T_p$ we potentially expect finite resistance in the {\it c}-direction
and zero resistance within the planes. This has turned to be difficult to
find, both experimentally \cite{norejec} and in our simulations. The
resistance seems rather to go to zero at the same temperature $T_i$ in all
directions.\cite{nota1} This is due to the fact that vortex paths crossing
the sample for $T>T_p$ are pinned to the {\it ab}-planes as long as $T<T_i$,
preventing their movement (and thus dissipation). However, it is worth
noting that some other experimental measurements of coherence (AC
magnetization) indicate\cite{exper1} that in fact, {\it c}-axis coherence is
lost at lower temperatures than in-plane coherence for the case of
Bi$_2$Sr$_2$CaCu$_2$O.

The transitions observed in our simulations are of different character, and
we want to discuss the point a bit further. The first order transition is
the easiest to characterize numerically. Although we do not show all the
results, we observed that when the resistivity has a jump other indicators
point to a first order phase transition, among them the existence of
hysteresis in the resistivity curves upon heating and cooling, and the fact
that right at the transition temperature, the energy histogram of the system
has two peaks, indicating two coexisting phases with an energy barrier
separating them.\cite{teoria1ord,huse,mingo1} The continuous transitions are
more difficult to characterize. The transition at $T_i$ {\it is not} a phase
transition in our model. In fact, it is a crossover due to thermal deppining
of rather independent vortices.\cite{jb1,jb3} However, in a real sample it
may correspond to the vortex glass transition, depending on the strength of
the disorder.\cite{probrae} When $T_p>T_i$, we have previously characterized
the transition at $T_p$ as a percolation phase transition of the vortex
structure perpendicularly to the applied field.\cite{jb3} In the
thermodynamic limit for $L_{ab}$ ($L_{ab}\rightarrow \infty $) the system
does not have any vortex line running perpendicularly to the applied field
for $T<T_p$, whereas for $T>T_p$ these paths extend all over the {\it ab}
plane with probability one. In References [14,12] we showed numerical
evidence suggesting that this transition is a second order phase transition
and gave its critical exponents as found from simulations.

\section{Summary and conclusions}

In this paper we presented numerical evidence that supports an
anisotropy-disorder phase diagram of the vortex structure of high-$%
T_c^{\prime }s$ with the following characteristics: For clean samples the
vortex lattice melts through a first order phase transition for a wide range
of anisotropies. When disorder is included the behavior of the system is
strongly dependent of the anisotropy. For low anisotropies the in-plane
coherence is lost at a temperature $T_i$ lower than the temperature $T_p$ at
which inter-plane coherence is lost, and a zone of disentangled vortex lines
is observed for $T_i<T<T_p$. For highly anisotropic samples the
superconducting coherence as deduced from simulations of the resistivity is
lost at the same temperature $T_i$ within the planes and perpendicularly to
the planes. However, in this case the vortex structure percolates at a
temperature $T_p$ well below $T_i$. In this case the system for $T_p<T<T_i$
is in an `entangled solid' phase. These features are also obtained from an
estimation of the free energy of the system which is mainly based on a
proposal for the entropy of the system (Eq. (\ref{entropia})). We showed
that the magnetic field-temperature behavior of the system can be deduced
from results obtained from a fixed magnetic field provided the anisotropy
and disorder present in the system are properly rescaled.

Our results present in a unified way, different characteristics of the
vortex structure that had been previously found in partial studies. The
analysis is in agreement with a variety of experiments performed on
different materials with a broad range of parameters such as disorder,
anisotropy and magnetic field. It could prove to be useful to find a more
solid base of our proposal for the free energy of the system -that we showed
is qualitative good- in order to obtain more detailed analytical results.

\section{Acknowledgments}

We thank helpful discussions with D. L\'{o}pez, E. F. Righi and F. de la
Cruz. E.A.J. acknowledges financial support by CONICET. C.A.B. is partially
supported by CONICET.

\end{multicols}

\end{document}